\documentclass[review]{elsarticle}
\usepackage{float}
\usepackage{geometry}
\usepackage{epstopdf}
\usepackage{url}
\usepackage{hyperref}
\usepackage{color}
\usepackage{amsmath}

\journal{Physics Letters A}

\begin{document}

\begin{frontmatter}

\title{Comparative analysis of the original and amplitude permutations}

\author[mainaddress]{Wenpo Yao\corref{mycorrespondingauthor}}
\cortext[mycorrespondingauthor]{Corresponding author}
\ead{yaowp@njupt.edu.cn}
\author[Secaddress]{Wenli Yao}
\author[mainaddress]{Jun Wang}

\address[mainaddress]{School of Geographic and Biologic Information, Nanjing University of Posts and Telecommunications, Nanjing 210023, China}
\address[Secaddress]{Department of hydraulic engineering, School of civil engineering, Tsinghua University, Beijing 100084, China}

\begin{abstract}
The original and amplitude permutations are two basic ordinal patterns; however, their relationship has received little attention. This paper compares the original and amplitude permutations used to characterize vector structures. To accurately convey the vector structure, we modify indexes of equal values in the permutations to be the same ones in each group of equalities. Comparative analysis suggests that the amplitude permutation, comprising the positions of the original values in the reordered vector, directly reflects the vector's temporal structure, whereas the original permutation, consisting of the indexes of reorganized values in the original vector, conveys the structural pattern of the reorganized vector. Moreover, we clarify the association of the original and amplitude permutations with time- and amplitude-symmetric vectors, thus contributing to the fields of symbolic analysis, topological data analysis, and so on.
\end{abstract}


\end{frontmatter}


\section{Introduction}
Rank-based techniques have long been used in time-series analysis \cite{Hallin1994}. Since the introduction of permutation entropy \cite{Bandt2002P}, permutation analysis has gained increasing popularity as a means of simplifying a time series and extracting its dynamical features. The coarse-grained ordinal method maps the time series into a sequence of permutations on the basis of a comparison of neighboring values. Ordinal approaches inherit causal information about the dynamical processes and have been widely used in physics, mathematics, engineering, and biomedicine \cite{Pessa2020,Zanin2021,Bandt2020,Echegoyen2019,Zanin2012,Amigo2015,Bandt2016,Yao2020CNS,Yao2020ND}.

Despite containing inherent information about the temporal structure, the original permutation (OrP) \cite{Bandt2002P} is not a direct reflection of the amplitude structure of vectors because it comprises the indexes of reorganized values in the original vector. This limitation is irrelevant if the order pattern is simply used as a label of the vector, such as in permutation entropy, synchronization or networks \cite{Bandt2002P,Pessa2020,Zanin2021,Bandt2020,Echegoyen2019,Zanin2012,Amigo2015,Bandt2016}. However, it might yield misleading results or even conceptual errors if the OrP is used as a temporal replacement of the vector, such as in time irreversibility \cite{Yao2020CNS,Yao2020ND}. For example, the present authors used to \textit{wrongly employ symmetric OrPs as alternatives to symmetric vectors} for an investigation of quantitative time irreversibility \cite{Yao2019Ys}. This approach is incorrect because the original symmetric permutations are not always equivalent to the permutations of symmetric vectors \cite{Yao2020CNS,Yao2020ND}. In other literature \cite{Zanin2021,Bandt2020,Echegoyen2019,Zanin2012,Amigo2015,Amigo2010,Zanin2018}, the position of the sequence elements in the reorganized series (i.e., the relationship between amplitude sizes) has been adopted as a faithful reflection of the temporal structure of vectors. We refer to this alternative ordinal scheme as the amplitude permutation (AmP).

Intuitively, the AmP would appear to be better than the OrP for characterizing temporal structures and in permutation analysis. However, as the OrP provides an indirect expression of the spatial structure, it is useful in various applications. For example, besides scientific research \cite{Bandt2002P,Pessa2020,Yao2019Ys}, MATLAB's `sort' \cite{sort} and PYTHON's 'numpy.argsort' \cite{argsort} functions both return the arrangement type of vector elements in the form of the OrP. Both the OrP and AmP convey the temporal structures of vectors. They are closely related to each other, and can even be the same under certain conditions. However, the association between the OrP and AmP and the implications of their relationship remain unknown.

In this paper, we present a comparative analysis of the OrP and AmP using some specific example vectors, particularly those with equal values. Our findings highlight that the AmP directly reflects the vector temporal structure and clarify the relationships between the OrP and AmP, as well as demonstrating the links between the basic permutations and vectors for related symbolic or
topological data analysis.

\section{Basic original and amplitude permutations}
Let us first introduce the original and amplitude ordinal schemes. Given the time series $X(t)=\{x(1),\ldots,x(t),\ldots,x(L)\}$, we construct multi-dimensional vectors with dimension $m$ and delay $\tau$ as follows:

\begin{eqnarray}
	\label{eq1}
	X_{m}^{\tau}(i)=\{ x(i),x(i+\tau),\ldots,x(i+(m-1)\tau)\} .
\end{eqnarray}

Regarding the space vector $X(i)=\{x(i_{1}),\ldots,x(i_{i}),\ldots,x(i_{m})\}$ in $X_{m}^{\tau}(i)$, we rearrange the elements in ascending order of their relative values to give $X(j)=\{x(j_{1}),\ldots,x(j_{j}),\ldots,x(j_{m})\}$ such that $x(j_{1})< \ldots < x(j_{j})< \ldots <x(j_{m})$. Note that the index $i$, ranging from 1 to $m$, denotes the position of the element in the original vector $X(i)$, and the index $j$ represents the position of the element in the ascending series $X(j)$.

We join the indexes of the original vector $X(i)$ and reordered vector $X(j)$ to generate a new form of vector, $X(i,j)$. To conduct the OrP, we rewrite $X(j)$ as follows:

\begin{eqnarray}
	\label{eq2}
	X(i,j)=\{x(i,1),x(i,2),\ldots,x(i,j-1),x(i,j),x(i,j+1),\ldots,x(i,m-1),x(i,m)\} ,
\end{eqnarray}
where $j$ ranges from 1 to $m$, and $i$ is the position of $x(j)$ in the original vector $X(i)$.

Next, we generate the series of $i$ values for $X(i,j)$ as $\pi_{i}=(i_{1},i_{2},\ldots,i_{j-1},i_{j},i_{j+1},\cdots, i_{m-1},i_{m})$, i.e., the OrP. This ordinal approach describes the positions of the reorganized values $x(j)$ in the original vector $X(i)$ \cite{Bandt2002P,Pessa2020,Yao2019Ys}. The OrP is widely used; for example, the `sort' function \cite{sort} in MATLAB and the 'argsort' function \cite{argsort} in PYTHON both return the ascending series as well as the sequence of original indexes of the reordered $x(j)$ in the series $X(i)$.

For the AmP, $X(i)$ is modified as follows:

\begin{eqnarray}
	\label{eq3}
	X(i,j)=\{x(1,j),x(2,j),\ldots,x(i-1,j),x(i,j),x(i+1,j),\ldots,x(m-1,j),x(m,j)\} ,
\end{eqnarray}
where $i$ ranges from 1 to $m$, and $j$ is the position of $x(i)$ in the reordered vector $X(j)$.

The AmP is the series of $j$ values, i.e., $\pi_{j}=(j_{1},j_{2},\ldots,j_{i-1},j_{i},j_{i+1},\cdots, j_{m-1},j_{m})$. These values clarify the positions of the original elements $x(i)$ in the reorganized series $X(j)$, i.e., the direct temporal structure of $X(i)$ \cite{Zanin2021,Bandt2020,Echegoyen2019,Zanin2012,Amigo2015,Amigo2010,Zanin2018}.

Overall, both the OrP and AmP specify how the elements in the original vector $X(i)$ have been reorganized. The difference is as follows: the vector $X(i,j)$ of the OrP in Eq.~(\ref{eq2}) is the same as the reordered vector $X(j)$, and the OrP consists of the indexes of $x(i)$ in $X(i)$ with respect to $X(j)$; in contrast, the vector $X(i,j)$ of the AmP in Eq.~(\ref{eq3}) is the same as the original vector $X(i)$, and the AmP comprises the indexes of $x(j)$ in the reordered vector $X(j)$ according to $X(i)$.

\section{Comparison of the OrP and AmP}
Let us demonstrate the differences and associations between the OrP and AmP using some specific example vectors.

\subsection{OrP and AmP of symmetric vectors}
\begin{figure}[htb]
	\centering
	\includegraphics{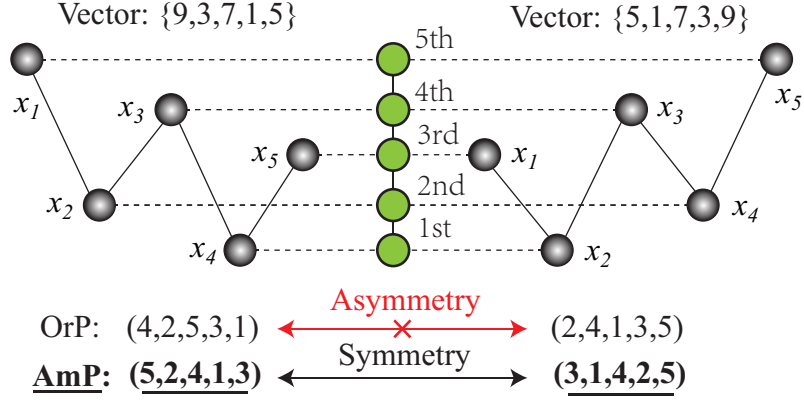}
	\caption{A pair of symmetric vectors and their OrPs and AmPs. The five green elements correspond to the amplitude positions of the elements in each vector. The AmPs (bold and underlined) of these symmetric vectors are symmetric.}
	\label{fig1}
\end{figure}

We consider the symmetric vectors \{9,3,7,1,5\} and \{5,1,7,3,9\} as examples and construct their OrPs and AmPs comparatively as shown in Fig.~\ref{fig1}.

When the elements are rearranged in ascending order, these vectors are reorganized to \{1,3,5,7,9\}. Taking the vector \{9,3,7,1,5\} in Fig.~\ref{fig1} as an example, the consecutive positions of the values in the rearranged vector \{1,3,5,7,9\} give the AmP, i.e., (5,2,4,1,3). The first value `5' in the AmP means that the first element $x_{1}$ ranks 5th according to amplitude. The original index of the values in the vector \{9,3,7,1,5\} with respect to the rearranged vector \{1,3,5,7,9\} is the OrP, i.e., (4,2,5,3,1). The first value `4' in the OrP implies that the first element in the reordered vector \{1,3,5,7,9\} is the 4th element $x_{4}$ in the original \{9,3,7,1,5\}.

The AmP is directly in line with the amplitude values of the original vectors, whereas the OrP is relevant to the positions of ascending values in the original vector. Therefore, the two AmPs of the symmetric vectors are symmetric, whereas the pair of OrPs are not symmetric.

\subsection{OrP and AmP of equal-values vectors}
Equal values play an important role in the ordinal scheme and have significant effects on permutation analysis \cite{Bian2012,Zunino2017,David2018,Yao2019E}, which has not been paid enough attention. In some signals, equal states are not rare; for example, the heart rates of patients with congestive heart failure might have about 46\% neighboring equal heartbeats \cite{Yao2019E}. The equal values change the construction and probability distributions of ordinal patterns, thus yielding significantly different and even contradictory outcomes \cite{Zunino2017,David2018,Yao2019E,Yao2020APL}. Moreover, in time irreversibility, vectors containing equal states might be self-symmetric (i.e., the same as their symmetric form), and this has a specific physical implication, i.e., time reversibility or temporal symmetry \cite{Yao2020CNS,Yao2019Ys,Yao2020ND,Yao2019E}. Currently, equal values are numerically broken by adding small random perturbations or just be arranged following their orders of occurrence \cite{Bandt2002P,Yao2019Ys,Amigo2010,sort,argsort,Herrera2020,Martinez2018}, which is not appropriate.

To deal with equal values in the generation of permutations, we rewrite their corresponding indexes into equal forms. We rearrange the equal values in neighboring orders according to their orders of occurrence as $\cdots<x(i_{1},j_{1})=x(i_{2},j_{2})< \cdots <x(i_{4},j_{4})=x(i_{5},j_{5})=x(i_{6},j_{6})<\cdots$. Next, we modify the indexes of the equal values to be the same as those in their corresponding groups, e.g., by the smallest index as $\cdots<x(i_{1},j_{1})=x(i_{1},j_{1})< \cdots <x(i_{4},j_{4})=x(i_{4},j_{4})=x(i_{4},j_{4})<\cdots$, as suggested by Bian et al. \cite{Bian2012}, or by the largest index as $\cdots<x(i_{2},j_{2})=x(i_{2},j_{2})< \cdots <x(i_{6},j_{6})=x(i_{6},j_{6})=x(i_{6},j_{6})<\cdots$. The OrP $\pi_{i}=(\ldots,i_{1},i_{2},\ldots,i_{4},i_{5},i_{6},\cdots)$ can be modified by the smallest indexes as $\pi_{i}=\{\ldots,i_{1},i_{1},\ldots,i_{4},i_{4},i_{4},\cdots\}$ or by the largest ones as $\pi_{i}=\{\ldots,i_{2},i_{2},\ldots,i_{6},i_{6},i_{6},\cdots\}$. Additionally, the equal-values AmP can be derived as $\pi_{j}=(\ldots,j_{1},\ldots,j_{1},\ldots,j_{4},\ldots,j_{4},\cdots\,j_{4},\ldots)$ or $\pi_{j}=(\ldots,j_{2},\ldots,j_{2},\ldots,j_{6},\ldots,j_{6},\cdots\,j_{6},\ldots)$. Note that the indexes of the equal values in the OrP are in consecutive order, whereas those in the AmP might not be.

\begin{figure}[htb]
	\centering
	\includegraphics{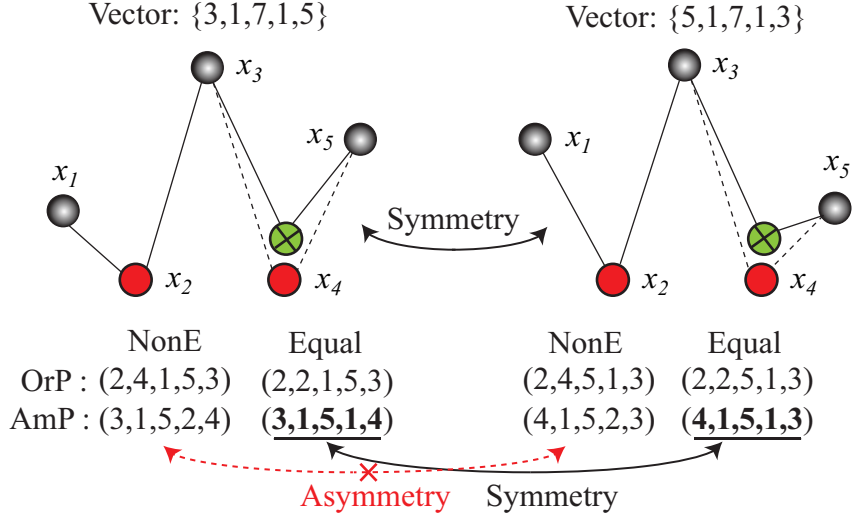}
	\caption{Symmetric vectors containing equal values and their permutations. The red elements are equal values, and the crossed green elements are alternatives according to their orders of occurrence. The indexes of equal values in the OrP and AmP are modified to be the smallest index '1' in the equal-values group. The order patterns under `NonE' refer to the original method without considering equal values, and those under `Equal' are equal-values permutations. Only the equal-values AmPs (bold and underlined) of these symmetric vectors are symmetric.}
	\label{fig2}
\end{figure}

Fig.~\ref{fig2} depicts the original and equal-value OrPs and AmPs of a pair of symmetric vectors containing two equal values, i.e., \{3,1,7,1,5\} and \{5,1,7,1,3\}. The double-equal values `1' in these vectors are not neighboring. By contrast, their indexes in the OrPs, namely the `NonE' (2,4,...) and the `Equal' (2,2,...) or (4,4,...), are neighboring. Therefore, the pairs of OrPs (2,2,1,5,3) and (2,2,5,1,3) in Fig.~\ref{fig2}, which could instead be (4,4,1,5,3) and (4,4,5,1,3), are not symmetric. The `NonE' AmPs of the symmetric vectors, i.e., (3,1,5,2,4) and (4,1,5,2,3), are also not symmetric. When we rewrite the AmPs according to the equal-value ordinal scheme, the `Equal' AmPs that are modified either by the smallest `1' to (3,1,5,1,4) and (4,1,5,1,3), as shown in Fig.~\ref{fig2}, or by the largest `2' to (3,2,5,2,4) and (4,2,5,2,3) are both symmetric. 

The equal-values permutation should be employed in ordinal approaches owing to its significant effects and physical implications \cite{Yao2020CNS,Yao2020ND,Yao2019Ys,Zunino2017,David2018,Yao2019E,Yao2020APL}. Moreover, the equal-values AmP faithfully reflects the structural information in vectors.

\subsection{OrP, AmP, and vector structure}
\begin{figure}[htb]
	\centering
	\includegraphics{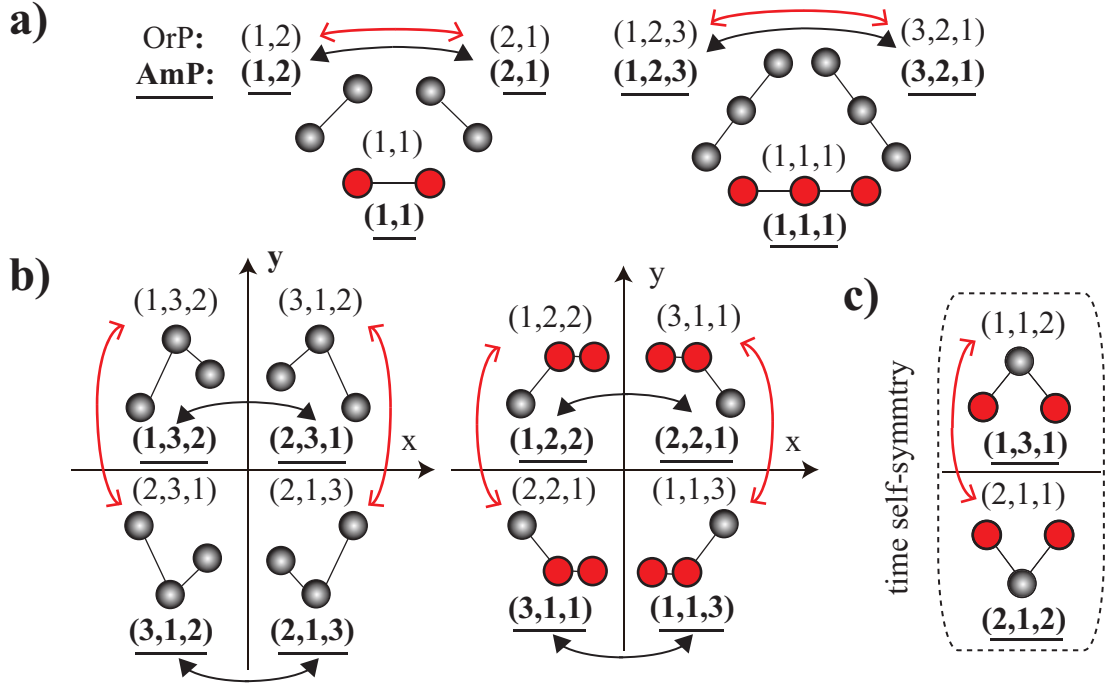}
	\caption{The OrPs and AmPs of vectors when $m$=2 and $m$=3. The red elements are equal values whose indexes are modified to be the smallest ones in their corresponding groups. The red arrows point to amplitude (x-axis) symmetric vectors and their symmetric OrPs, and the black arrows point to time (y-axis) symmetric vectors and their symmetric AmPs (bold and underlined). OrPs and AmPs of vectors in \textbf{a)} are always the same. Vectors and their OrPs and AmPs in \textbf{b)} are present in amplitude and time (say x- and y-axis) symmetric forms. \textbf{c)} shows a pair of time self-symmetric vectors containing double-equal values.}
	\label{fig3}
\end{figure}

The previous subsections suggest that the AmP is superior to the OrP, which is true in the amplitude-structure extraction. However, the OrP and AmP both reflect features of the spatial structure of vectors, although from different perspectives. Let us further demonstrate the association of the OrP, AmP, and vector structure from a comprehensive viewpoint, i.e., when $m$=2 and $m$=3, as shown in Fig.~\ref{fig3}.

All-up, all-down, and all-equal vectors always share the same OrPs and AmPs, as suggested by Fig.~\ref{fig3}\textbf{a)}. Moreover, if a vector has the same OrP and AmP, so does its central-symmetric form, e.g., the vectors in the top-left and bottom-right parts of the two group vectors in Fig.~\ref{fig3}\textbf{b)} and those in Fig.~\ref{fig3}\textbf{a)}. In Fig.~\ref{fig3}\textbf{b)}, the AmPs of symmetric vectors (e.g., y-axis or time symmetric), with or without equal values, are all symmetric. More importantly, the AmPs of the two time self-symmetric vectors in Fig.~\ref{fig3}\textbf{c)} are also self-symmetric. Therefore, the AmP reliably reflects the vector temporal structures. This feature is particularly relevant for time irreversibility because symmetric forms in the AmPs directly illustrate time symmetry and reversibility, avoiding the possible conceptual errors caused by the OrPs \cite{Yao2020CNS,Yao2019Ys}.

Generally speaking, our conventional wisdom about symmetry concerns time symmetry, i.e., y-axis symmetry. Under this condition, the AmPs of symmetric vectors are symmetric, as suggested by the vectors and AmPs in Figs.~\ref{fig1} and \ref{fig2} and those linked by black solid arrows in Fig.~\ref{fig3}. However, if we consider amplitude symmetry, i.e., the x-axis symmetry, the OrPs of symmetric vectors, connected by red arrows in Figs.~\ref{fig3}\textbf{b)} and \ref{fig3}\textbf{c)}, are also symmetric.

The corresponding values in time- or y-axis symmetric vectors have the same amplitude, and their AmPs are the same. By contrast, those in amplitude- or x-axis symmetric vectors have the same location index in the reorganized vector, and their OrPs are the same. Note that if the orders of all the amplitudes and indexes of the values in the reordered vector are the same, the OrP and AmP of this vector will be the same, e.g., the vectors in Fig.~\ref{fig3}\textbf{a)} and those in the top-left and bottom-right parts of Fig.~\ref{fig3}\textbf{c)}. The relationship between permutations and vectors is the same if we use the largest indexes for equal values in their corresponding groups. This association between the OrP and AmP suggests that the AmP can help address the problem of misleading permutations in time irreversibility, whereas the OrP is appropriate for simplifying quantitative amplitude irreversibility \cite{Yao2020CNS}.

Comparative analysis of the two ordinal patterns suggests that the AmP directly reflects the temporal structure of vectors, whereas the OrP implies the structural patterns about how the vectors are reorganized. This difference has no effect on the permutation entropy, causality detection or networked analysis \cite{Bandt2002P,Pessa2020,Zanin2021,Bandt2020,Echegoyen2019,Zanin2012,Amigo2015,Bandt2016}, because they are just labels of different vectors. If the ordinal pattern is employed as an alternative of the vector structure, the OrP and AmP should be selected accordingly to avoid incorrect reports, such as in time irreversibility \cite{Yao2020CNS}.

\section{Conclusions}
To conclude, the original permutation specifies the positions of reordered vector values in the original vector, whereas the amplitude permutation indicates the positions of original vector elements in the reordered vector. The amplitude permutation directly reflects the temporal structure of the original vector. The amplitude permutations of temporal-symmetric vectors are symmetric, and the original permutations of amplitude-symmetric vectors are symmetric. More importantly, the equal-values ordinal scheme is required no matter whether the distribution of equal values is large or not. The association identified in this paper between the original and amplitude permutations will lead to further theoretical and experimental demonstrations in areas such as symbolic time series analysis, topological data analysis, and numerical simulation.

\nocite{*}

\bibliography{mybibfile}

\begin{thebibliography}{10}
\expandafter\ifx\csname url\endcsname\relax
  \def\url#1{\texttt{#1}}\fi
\expandafter\ifx\csname urlprefix\endcsname\relax\def\urlprefix{URL }\fi
\expandafter\ifx\csname href\endcsname\relax
  \def\href#1#2{#2} \def\path#1{#1}\fi

\bibitem{Hallin1994}
M.~Hallin, M.~L. Puri, Aligned rank tests for linear models with autocorrelated
  error terms, Journal of Multivariate Analysis 50~(2) (1994) 175--237.

\bibitem{Bandt2002P}
C.~Bandt, B.~Pompe, Permutation entropy: a natural complexity measure for time
  series, Physical Review Letters 88~(17) (2002) 174102.

\bibitem{Pessa2020}
A.~A. Pessa, H.~V. Ribeiro, Mapping images into ordinal networks, Physical
  Review E 102~(5) (2020) 052312.

\bibitem{Zanin2021}
M.~Zanin, F.~Olivares, Ordinal patterns-based methodologies for distinguishing
  chaos from noise in discrete time series, Communications Physics 4~(1) (2021)
  1--14.

\bibitem{Bandt2020}
C.~Bandt, Order patterns, their variation and change points in financial time
  series and brownian motion, Statistical Papers 61~(5) (2020) 1565--1588.

\bibitem{Echegoyen2019}
I.~Echegoyen, V.~Vera-Ávila, R.~Sevilla-Escoboza, J.~H. Martínez, J.~M.
  Buldú, Ordinal synchronization: Using ordinal patterns to capture
  interdependencies between time series, Chaos, Solitons \& Fractals 119 (2019)
  8--18.

\bibitem{Zanin2012}
M.~Zanin, L.~Zunino, O.~A. Rosso, D.~Papo, Permutation entropy and its main
  biomedical and econophysics applications: A review, Entropy 14~(8) (2012)
  1553--1577.

\bibitem{Amigo2015}
J.~M. Amigo, K.~Keller, V.~A. Unakafova, Ordinal symbolic analysis and its
  application to biomedical recordings, Philosophical Transactions of the Royal
  Society A: Mathematical, Physical and Engineering Sciences 373~(2034) (2015)
  20140091.

\bibitem{Bandt2016}
C.~Bandt, Permutation Entropy and Order Patterns in Long Time Series, Springer,
  2016, pp. 61--73.

\bibitem{Yao2020CNS}
W.~P. Yao, J.~Wang, M.~Perc, W.~L. Yao, J.~Dai, D.~Guo, D.~Yao, Time
  irreversibility and amplitude irreversibility measures for nonequilibrium
  processes, Communications in Nonlinear Science and Numerical Simulation 96
  (2021) 105688.

\bibitem{Yao2020ND}
W.~Yao, J.~Dai, M.~Perc, J.~Wang, D.~Yao, D.~Guo, Permutation-based time
  irreversibility in epileptic electroencephalograms, Nonlinear Dynamics
  100~(1) (2020) 907--919.

\bibitem{Yao2019Ys}
W.~P. Yao, W.~L. Yao, J.~Wang, J.~Dai, Quantifying time irreversibility using
  probabilistic differences between symmetric permutations, Physics Letters A
  383~(8) (2019) 738--743.

\bibitem{Amigo2010}
J.~M. Amigo, Permutation complexity in dynamical systems: ordinal patterns,
  permutation entropy and all that, Springer Science \& Business Media, 2010.

\bibitem{Zanin2018}
M.~Zanin, A.~Rodríguez-González, E.~Menasalvas~Ruiz, D.~Papo, Assessing time
  series reversibility through permutation patterns, Entropy 20~(9) (2018) 665.

\bibitem{sort}
MathWorks, Sort array elements, [EB/OL],
  \url{https://www.mathworks.com/help/matlab/ref/sort.html}.

\bibitem{argsort}
The-NumPy-community, numpy.argsort, [EB/OL],
  \url{https://numpy.org/doc/stable/reference/generated/numpy.argsort.html}.

\bibitem{Bian2012}
C.~Bian, C.~Qin, Q.~D. Ma, Q.~Shen, Modified permutation-entropy analysis of
  heartbeat dynamics, Physical Review E 85~(2 Pt 1) (2012) 021906.

\bibitem{Zunino2017}
L.~Zunino, F.~Olivares, F.~Scholkmann, O.~A. Rosso, Permutation entropy based
  time series analysis: Equalities in the input signal can lead to false
  conclusions, Physics Letters A 381~(22) (2017) 1883--1892.

\bibitem{David2018}
D.~Cuesta–Frau, M.~Varela–Entrecanales, A.~Molina–Picó, B.~Vargas,
  Patterns with equal values in permutation entropy: Do they really matter for
  biosignal classification?, Complexity 2018 (2018).

\bibitem{Yao2019E}
W.~Yao, W.~Yao, J.~Wang, Equal heartbeat intervals and their effects on the
  nonlinearity of permutation-based time irreversibility in heart rate, Physics
  Letters A 383~(15) (2019) 1764--1771.

\bibitem{Yao2020APL}
W.~Yao, W.~Yao, D.~Yao, D.~Guo, J.~Wang, Shannon entropy and quantitative time
  irreversibility for different and even contradictory aspects of complex
  systems, Applied Physics Letters 116~(1) (2020) 014101.

\bibitem{Herrera2020}
J.~L. Herrera-Diestra, J.~M. Buldú, M.~Chavez, J.~H. Martínez, Using symbolic
  networks to analyse dynamical properties of disease outbreaks, Proceedings of
  the Royal Society A 476~(2236) (2020) 20190777.

\bibitem{Martinez2018}
J.~H. Martinez, J.~L. Herrera-Diestra, M.~Chavez, Detection of time
  reversibility in time series by ordinal patterns analysis, Chaos: An
  Interdisciplinary Journal of Nonlinear Science 28~(12) (2018) 123111.

\end{thebibliography}

\end{document}